\documentclass[useAMS,usenatbib]{mn2e}
\usepackage{times}
\usepackage{verbatim}
\usepackage{amsmath}
\usepackage{graphicx}
\usepackage{color}
\usepackage{aas_macros}
\usepackage{graphicx}
\usepackage{epsf}
\usepackage{graphics}
\usepackage{rotating}
\usepackage{amssymb}   
\usepackage{setspace}  
\usepackage{lscape}

\newcommand{\etal}{{\it et al.}}
\newcommand{\be}{\begin{equation}}
\newcommand{\ee}{\end{equation}}

\defcitealias{graves09bb}{GFS}
\defcitealias{magoulas11x}{MSC}

\title{The 6dF Galaxy Survey: Stellar Population Trends Across and Through the Fundamental Plane}

\author[Springob et al.]
{Christopher M. Springob$^{1}$, Christina Magoulas$^{2}$, Rob Proctor$^{3}$, Matthew Colless$^{1}$,
\newauthor
D. Heath Jones$^{1,4}$, Chiaki Kobayashi$^{5}$, Lachlan Campbell$^{5,6}$, John Lucey$^{7}$,
\newauthor
\& Jeremy Mould$^{8,9}$ \\
$^1$Australian Astronomical Observatory, PO Box 296, Epping, NSW 1710, Australia\\
$^2$School of Physics, University of Melbourne, Parkville, VIC 3010, Australia\\
$^3$Universidade de Sao Paulo, IAG, Rua do Mato 1226, S‹o Paulo 05508-900, Brazil\\
$^4$School of Physics, Monash University, Clayton, VIC 3800, Australia\\
$^5$Research School of Astronomy \& Astrophysics, The Australian National University, Cotter Rd., Weston ACT 2611, Australia\\
$^6$Department of Physics \& Astronomy, University of Western Kentucky, Bowling Green, KY 42102-3576 USA\\
$^7$Department of Physics, University of Durham, Durham DH1 3LE\\
$^8$Centre for Astrophysics and Supercomputing, Swinburne University, Hawthorn, VIC 3122, Australia\\
$^9$ARC Centre of Excellence for All-sky Astrophysics (CAASTRO)\\
}

\begin{document}

\maketitle

\begin{abstract}

We present results from an analysis of stellar population parameters for 7132 galaxies in the 6dFGS Fundamental Plane (FP) sample.  We bin the galaxies along the axes, $v_1$, $v_2$, and $v_3$, of the tri-variate Gaussian to which we have fit the galaxy distribution in effective radius, surface brightness, and central velocity dispersion (FP space), and compute median values of stellar age, [Fe/H], [Z/H], and [$\alpha$/Fe].  We determine the directions of the vectors in FP space along which each of the binned stellar population parameters vary most strongly.  In contrast to previous work, we find stellar population trends not just with velocity dispersion and FP residual, but with radius and surface brightness as well.  The most remarkable finding is that the stellar population parameters vary through the plane ($v_1$ direction) and across the plane ($v_3$ direction), but show no variation at all along the plane ($v_2$ direction).  The $v_2$ direction in FP space roughly corresponds to `luminosity density'.  We interpret a galaxy's position along this vector as being closely tied to its merger history, such that early-type galaxies with lower luminosity density are more likely to have undergone major mergers.  This conclusion is reinforced by an examination of the simulations of Kobayashi (2005), which show clear trends of merger history with $v_2$.
\end{abstract}


\section{Introduction}

Early-type galaxies are known to lie along a plane in the 3-dimensional (3D) parameter space whose axes are $r$=log($R_e$), $s$=log($\sigma$), and $i$=log($I_e$), where $R_e$, $\sigma$, and $I_e$ represent effective radius, central velocity dispersion, and effective surface brightness respectively.  This is commonly referred to as the Fundamental Plane (hereafter FP; \citealt{dressler87a,djorgovski87x}).  The plane can be expressed in the form

\be
r = as + bi +c
\ee

\noindent where $a$, $b$, and $c$ are observationally derived constants.  In the case where all early-type galaxies follow the virial theorem and have a constant mass-to-light ratio, $a = 2$ and $b = -1$.  In contrast to this theoretical expectation, the observed values are found to be in the range $1.2 < a < 1.6$ and $-0.90 < b < -0.75$ across a wide range of optical and near infrared wavebands (e.g., \citealt{lucey91x}, \citealt{pahre98x}, \citealt{hyde09bb}, \citealt{labarbera10aa}).  This contrast may be due at least in part to stellar population variations \citep{cappellari06x}, though it has been argued that nonhomology must also contribute to the tilt (see, e.g., \citealt{trujillo04}).

Stellar population variations lead to changes in the mass-to-light ratio, and may do so in ways that are correlated with FP parameters, leading to tilts of the FP.  Such variations may also introduce additional scatter in the relation.  Understanding these correlations and scatter may illuminate the origins of the FP and the formation of early-type galaxies.  It may also lead to a means of improving the accuracy and precision of the FP distance estimator.

Several authors have investigated correlations between FP parameters and stellar population (hereafter SP) parameters.  \citet{nelan05x}, \citet{thomas05a}, and \citet{smith07bb} all found strong correlations between $\sigma$ and several different SP parameters.  \citet{terlevich02aa} found that [Mg/Fe] increases with luminosity.  \citet{forbes98x}, \citet{reda05}, and \citet{gargiulo09a}, among others, found correlations between SP parameters and residuals from the FP.  \citet{labarbera10bb} found correlations between SP gradients and $\sigma$, stellar mass, and dynamical mass.  SP trends with radius and surface brightness individually, however, remained largely unexamined until recently.

In their series of four papers \citep{graves09aa,graves09bb,graves10x,graves10y}, Graves, Faber, \& Schiavon took the analysis of SP trends in FP space a step further than the earlier studies, by considering SP trends along all three dimensions of FP space.  They investigated SP parameter trends with radius, velocity dispersion, and surface brightness residuals from the FP for galaxies in the Sloan Digital Sky Survey Data Release 6 (\citealt{adelmanmccarthy08x}; hereafter SDSS), and found clear trends of several SP parameters with velocity dispersion, but much weaker trends with radius and surface brightness.

The authors found that while each of the SP parameters they studied increases with increasing velocity dispersion, there is little correlation between any SP parameter and radius.  In their Paper II (\citealt{graves09bb}; hereafter GFS), they hypothesized that galaxies with similar physical properties but different merger histories may vary widely in radius (e.g., \citealt{robertson06a}).  The fact that the SP parameters are insensitive to radius can thus be understood as an indication that SP parameters are determined independently of merger history.  One drawback of the Graves, Faber, \& Schiavon analysis, however, is that the authors bin galaxies along directions that are not orthogonal in FP space, which may potentially cause or hide correlations between SP and FP parameters.

In this paper, we investigate SP trends in FP space using data from the Six-degree Field Galaxy Survey (6dFGS).  6dFGS is a near infrared and optically selected dual redshift/peculiar velocity survey, with redshifts for over 125,000 galaxies in the southern hemisphere (\citealt{jones04,jones09x}).  9572 of those galaxies are included in our `velocity sample' (hereafter, 6dFGSv), which is described in Campbell \etal ~(in prep.).  For each of the galaxies in 6dFGSv, we have redshifts and velocity dispersions derived from 6dFGS, along with J-, H-, and K-band radii and surface brightnesses derived from 2MASS photometry.  We have fit the FP to this sample as described by Magoulas, Springob, \& Colless \etal ~(submitted, hereafter \citetalias{magoulas11x}).

Our ultimate aim is to derive distances and peculiar velocities for these galaxies, which will be used to characterize the local galaxy peculiar velocity field and provide constraints on cosmological models.  However, we would first like to explore whether the FP relation's utility as a redshift independent distance indicator can be improved by accounting for SP variations in the plane.  We would also like to gain a better understanding of both what SP trends mean for the star formation history in early-type galaxies, and what the distribution of galaxies in FP space means for both stellar and dynamical evolution.

To this end, we have derived values of four SP parameters (age, [Fe/H], [$\alpha$/Fe], and [Z/H]) for 7132 of the galaxies in 6dFGSv.  As described in Proctor \etal ~(in prep.), these were derived using Lick indices, following the procedure of \citet{proctor02}.  In Section 2, we discuss the dataset, and briefly describe the fitting of the Fundamental Plane, as well as the derivation of the SP parameters.  In Section 3, we present an analysis of the SP trends in FP space in which we bin the galaxies by position in FP space, calculate the median value of each SP parameter within each bin, and then calculate the gradient of the SP parameter variation, which gives us the vector in FP space along which the parameter varies.  In Section 4, we discuss the physical interpretation of the SP trends in the context of the 3D Gaussian distribution of galaxies in FP space.

\section{Data}

Complete details of the sample selection and data reduction are presented in Campbell \etal ~(in prep.) and \citetalias{magoulas11x}, but we summarize the relevant points here.  The 6dFGSv sample consists of all 6dFGS early-type galaxies that meet the following criteria:  spectral signal-to-noise ratio greater than 5, heliocentric redshift $z_{helio}<0.055$, log of velocity dispersion $s>2.05$ (in units of log[km/s]), and near infrared magnitude brighter than $m_j = 13.65$.  As explained in both Campbell \etal ~(in prep.) and \citetalias{magoulas11x}, ``early-type galaxies'' in this context includes spiral bulges in cases for which the bulge fills the 6dF fibre.  As \citetalias{magoulas11x} shows, spiral bulges follow essentially the same FP as early-type galaxies.  While there is a 0.04 dex  offset in the FP zeropoint between the sample of ellipticals and the sample of spiral bulges, there is actually no difference in the total thickness of the plane between the total sample and the sample of ellipticals.  Including spiral bulges has no impact on the thickness of the plane.

The apparent magnitudes used in this selection are taken from the Two Micron All-Sky Survey (2MASS) Extended Source Catalog \citep{jarrett00x}.  We have derived radii and surface brightnesses for three different overlapping samples of 6dFGSv, corresponding to J-, H-, and K-band, with slightly different limiting magnitudes.  Because the J-band sample offers photometric parameters with the smallest errors, that is our preferred passband, and we have fit the FP in J-band.  All photometric parameters used throughout this paper are in J-band.

We have derived velocity dispersions for each of these galaxies from their 6dFGS spectra.  We have also derived half-light radii and surface brightnesses from their 2MASS J-band photometric images.  Surface brightness is defined here as the average surface brightness interior to the half light radius.  The angular radii have been converted to physical radii using the redshift distance to the galaxy, as explained in \citetalias{magoulas11x}.  As mentioned in Section 1, we convert the physical radius, velocity dispersion, and surface brightness into logarithmic form, and write them as $r$, $s$, and $i$ respectively.

2872 of the galaxies are in groups or clusters, and we use the redshift distance of the group or cluster in such cases, where the group redshift is defined as the median redshift among all galaxies in the group.  This is done because the galaxies within a group will tend to be at roughly the same distance, and the systemic redshift distance of the group offers a better estimate of the distance to the galaxy than the individual galaxy redshift distance.  Further details on the grouping algorithm are found in \citetalias{magoulas11x}.

The initial morphological selection, described by Campbell \etal ~(in prep.), involves matching the galaxy spectrum to a sample of spectral templates, and retaining only galaxies whose spectra match those of early-type galaxies.  This selection allows for the inclusion of spirals, but is likely to do so only in cases for which the bulge fills the fibre.  \citetalias{magoulas11x} describes how we subsequently inspected images of each of the galaxies by eye, and eliminated cases with obvious problems, such as irregular morphologies.  Spirals were only eliminated in cases for which either the fibre included some contribution from the disk or the galaxy was edge on and included a visible dust lane.

\subsection{Fitting the Fundamental Plane}

\citetalias{magoulas11x} describes how we fit the FP using a maximum likelihood method that closely follows the procedure of EFAR \citep{colless01a,saglia01}.  The procedure is explained in detail in \citet{colless01a} Section 3.  It involves fitting the observed structural parameters to a 3D Gaussian in FP space.  That is, we assume that, when plotted in $r-s-i$ space, the galaxies follow a 3D Gaussian distribution, for which the two longest axes of the 3D Gaussian define the Fundamental Plane, and the shortest axis of the 3D Gaussian is orthogonal to the plane.  By construction, one of these axes has both an $r$ and an $i$ component, but no $s$ component.  The other two axes have components in all three dimensions.

Given this 3D Gaussian functional form, the probability density for the $i$th galaxy, $P(x_i)$, can then be computed according to \citet{colless01a} Equation 1, where $x_i$ is the $i$th galaxy's position in FP space relative to the axes of the 3D Gaussian.  Given each galaxy's position in FP space, we then fit the orientation of the 3D Gaussian's axes (and thus the zeropoint and slopes $a$ and $b$ of the FP), by finding the orientation of the Gaussian that maximizes the product of $P(x_i)$'s for every galaxy in the sample (the `likelihood').  (See Equations 2 and 6 in \citealt{colless01a}, and the corresponding explanation in Section 3 of the paper.)  The actual orientation of the FP that gives the maximum likelihood is found by searching the multi-dimensional parameter space with a non-derivative multi-dimensional optimization algorithm called BOBYQA (Bound Optimization BY Quadratic Approximation; \citealt{powell06a}).

The assumption of a 3D Gaussian distribution is motivated on purely empirical grounds.  There is no obvious theoretical reason for one to expect that galaxies would follow such a distribution in FP space.  However, when we fit to this model, the total likelihood of the fit (that is, the product of probability densities for every galaxy in the sample) is indistinguishable from that of mock catalogs that were generated under the assumption of 3D Gaussianity, suggesting a good fit.  Additionally, as \citetalias{magoulas11x} shows, the distributions of individual parameters closely matches those generated by the mock catalogs as well.  One might infer that the 3D Gaussian model implies that the luminosity function peaks at some intermediate luminosity, and symmetrically falls off at fainter luminosities.  This is not the case, however, as our selection cuts slice through FP space close to the center of the 3D Gaussian.  Thus, dwarf galaxies are largely excluded, and the shape of the distribution of galaxies at the fainter end of the FP is unobserved.

The best fit coefficients to the J-band FP $r = as + bi +c$ are $a=1.524 \pm 0.026$, $b=-0.885 \pm 0.008$, $c=-0.329 \pm 0.054$, where $r$, $s$, and $i$ are in units of log[kpc/h], log[km/s], and log[$L_{Sun}/pc^2$] respectively.  (Note: The `h' in kpc/h refers to the Hubble constant, in units of 100 km s$^{-1}$ Mpc$^{-1}$.  For the purpose of angular unit conversion, a flat cosmology of $\Omega_m=0.3$ and $\Omega_\Lambda =0.7$ is assumed, though the specifics of the assumed cosmology affect the FP fit very weakly.)

Following \citet{colless01a}, we refer to the three axes of the 3D Gaussian as $v_1$, $v_2$, and $v_3$.  The unit vectors along these axes are related to $r$, $s$, and $i$ by the FP slopes $a$ and $b$ (from Equation 1 of this paper) as follows:

\begin{align}
\mathbf{\hat{v}_{1}} &= \mathbf{\hat{r}} - a \mathbf{\hat{s}} - b \mathbf{\hat{i}},  \notag \\
\mathbf{\hat{v}_{2}} &= \mathbf{\hat{r}} + \mathbf{\hat{i}}/b, \\
\mathbf{\hat{v}_{3}} &= -\mathbf{\hat{r}}/b - (1 + b^{2})\mathbf{\hat{s}}/(ab) + \mathbf{\hat{i}} \notag
\end{align}

This closely follows \citet{colless01a}, though there is one small difference, as we define surface brightness in log[$L_{Sun}/pc^2$] units rather than magnitude units.  Given our measured values of $a=1.524$ and $b=-0.885$, we then have the transformation matrix between $r-s-i$ space and $v_1-v_2-v_3$ space, which we provide in Table 1.  That is, we provide the $r$-, $s$-, and $i$-components of unit vectors along $v_1$, $v_2$, and $v_3$ and vice versa.  (e.g., the unit vector along the $v_1$ direction has length 0.494 in the $r$-directions, while the $r$ unit vector correspondingly has length 0.494 in the $v_1$-direction.)  $v_1$ is the shortest axis of the 3D Gaussian, orthogonal to the plane.  It increases with increasing $r$ and $i$, but decreasing $s$.  $v_2$, the longest axis of the 3D Gaussian, increases with increasing $r$ and decreasing $i$, but has no $s$-component.  $v_3$ is the shorter of the two axes within the plane, and it increases with increasing $r$, $s$, and $i$.

As explained by \citetalias{magoulas11x}, the longest axis of the 3D Gaussian ($v_2$) has no $s$-component by construction.  We did perform one nine parameter fit (described in detail by \citetalias{magoulas11x}) in which we allowed $v_2$ to include an $s$-component.  The best fit value of the $s$-component of the $v_2$ unit vector is then -0.080.  And so, in the nine-parameter fit scenario, the central value of the transformation matrix in Table 1 becomes -0.080, and the other values in the matrix shift slightly so that every row and column is normalized to unity.  Since this represents a very small shift, we exclude the $s$-component of the $v_2$ vector for all other fits.

\begin{table}
\begin{tabular}{@{}lrrr@{}}
\hline
Axis of 3D Gaussian   &  $r$   &  $s$   &  $i$ \\
\hline
\hline
$v_1$ & 0.494 & -0.752 & 0.437 \\
$v_2$ & 0.663 & 0.000 & -0.749 \\
$v_3$ & 0.563 & 0.659 & 0.498 \\
\hline
\end{tabular}
\caption{Transformation matrix between $v_1-v_2-v_3$ and $r-s-i$.}
\end{table}

We also illustrate the directions of the $v_1$, $v_2$, and $v_3$ axes relative to the axes of physical parameters $r$, $s$, and $i$ in Figure 1\footnote{This plot is an interactive 3D visualization, undertaken with custom C-code and the S2PLOT graphics library \citep{barnes06x}.  Interactive 3D figures, which can be accessed by viewing the version of this paper found in the ancillary files on astro-ph with Adobe Reader Version 8.0 or higher, were created using the approach described in \citet{barnes08x}.}.  Also shown are the mass ($M$), luminosity ($L$), and mass-to-light ratio ($M/L$) directions, which we explain in more detail in Section 3.3.

\begin{figure}
\centering
\includegraphics[width=0.45\textwidth]{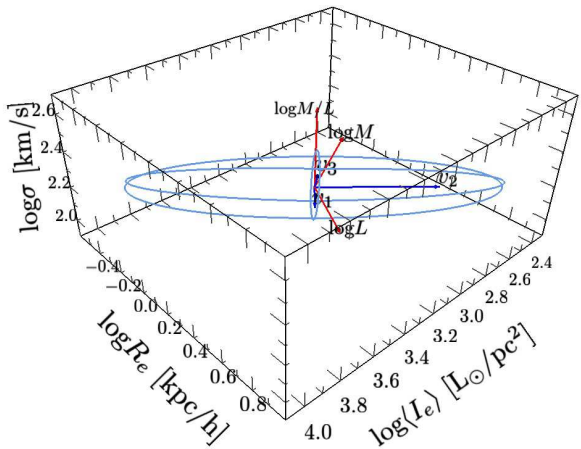}
\caption{A 3D representation of the directions of the axes of the 3D Gaussian to which the FP was fit, $v_1$, $v_2$, and $v_3$.  The $v_1$, $v_2$, and $v_3$ axes are given in blue, with the axis vectors drawn proportionally to the length of the Gaussian ($\sigma_1$, $\sigma_2$, and $\sigma_3$) along these three axes.  The wireframe ellipses also illustrate the length of the Gaussian along the three axes.  We note that $v_1$ is the shortest of the axes, and is thus orthogonal to the FP, while $v_2$ and $v_3$ are directions within the FP.  Also shown in red are the ``mass'', ``luminosity'', and ``mass/luminosity'' directions, as described in Section 3.3.  Readers using version 8.0 or higher of Adobe Reader can enable interactive, 3D views of this plot by mouse clicking on the version of this figure found in the ancillary files.  Once enabled, 3D mode allows the reader to rotate and zoom the view using the mouse.
\label{FIG1}}
\end{figure}

Following the convention of \citetalias{magoulas11x}, we refer to $v_1$, $v_2$, and $v_3$ as ``through the plane'', ``along the plane'', and ``across the plane'' respectively.

\subsection{Derivation of stellar population parameters}

The $\chi^2$-fitting procedure of \citet{proctor02} was used to measure the derived parameters: log(age), [Fe/H], [Z/H] and [E/Fe] (which we hereafter refer to as [$\alpha$/Fe], or the `$\alpha$' abundance ratio; see \citealt{thomas03bb} for details).  Briefly, the technique for deriving these parameters involves the simultaneous comparison of as many observed indices as possible to the model single stellar populations (SSPs) of \citet{korn05x}. The best fit is found by minimising the square of the deviations between observations and models in terms of the observational errors (i.e. $\chi^2$).

The rationale behind this approach is that, while all indices show some degeneracy with respect to each of the derived parameters, each index does contain {\it some} information regarding each parameter. In addition, such an approach should be relatively robust with respect to many problems that are commonly experienced in the measurement of spectral indices and their errors. These include poor flux calibration, poor sky subtraction, poorly constrained velocity dispersions, poor calibration to the Lick system and emission line contamination. This robustness is of particular importance in the analysis of large numbers of pipeline-reduced spectra such as those of the 6dFGS which cannot be accurately flux calibrated and so are not fully calibrated to the Lick system. The method is also relatively robust with respect to the uncertainties in the SSP models used in the interpretation of the measured indices; e.g. the second parameter effect in horizontal branch morphologies and the uncertainties associated with the Asymptotic Giant Branch. It was shown in \citet{proctor04y} and \citet{proctor05b} that the results derived using the $\chi^2$ technique are, indeed, significantly more reliable than those based on only a few indices.

Fitting was carried out using an iterative clipping procedure. Initially the data were fit and a 5$\sigma$ clip was applied. The data were then re-fit and a 3$\sigma$ clip was applied.  The fitting and 3$\sigma$ clipping were then iterated until no 3$\sigma$ outliers were found. Errors in the derived parameters were estimated using the Monte Carlo technique described in \citet{proctor08b}.  We note that, as described in \citet{proctor02}, the relationship between [Fe/H], [Z/H] and [$\alpha$/Fe] is fixed such that [Z/H]=[Fe/H]+0.942[$\alpha$/Fe].

A quality parameter was also defined as the sum of the integerised reduced-$\chi^2$ and the number of clipped indices.  Only data from galaxies with S/N per angstrom greater than 9 and quality parameter of 10 or lower are used in our analysis. As a result of the above procedures we measured the stellar population parameters in 7132 galaxies, each utilising 10 or more indices.

The distribution of each of the SP parameters is given in Figure 2.  The distribution of measurement errors on each of the SP parameters is given in Figure 3.  The values for each parameter for each individual galaxy will be presented in Proctor \etal ~(in prep.)

\begin{figure}
\centering
\includegraphics[width=0.45\textwidth]{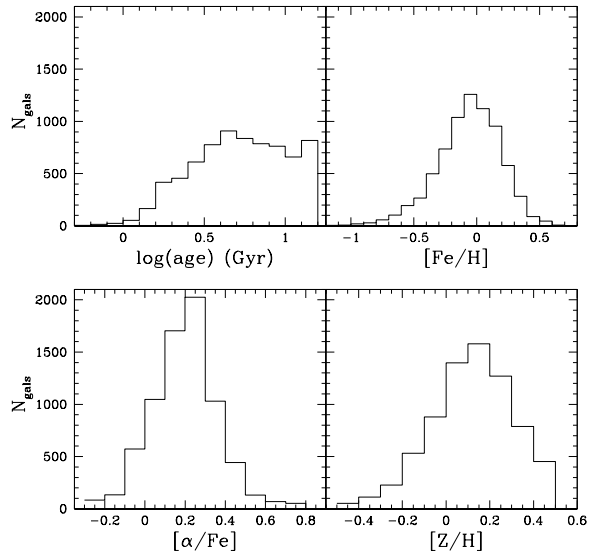}
\caption{Histograms of all four SP parameters.\label{FIG2}}
\end{figure}

\begin{figure}
\centering
\includegraphics[width=0.45\textwidth]{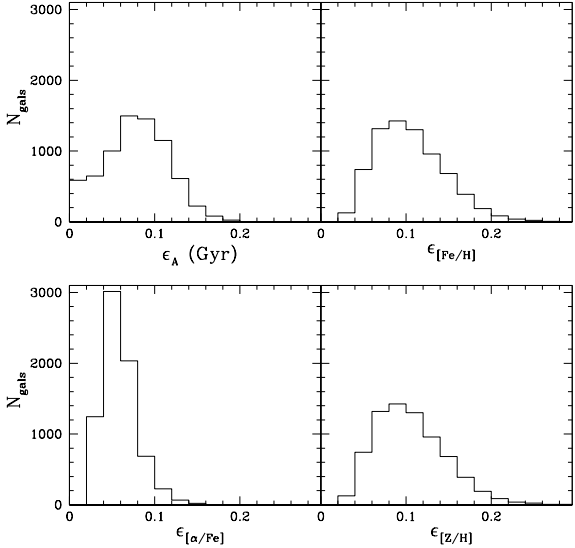}
\caption{Histograms of the statistical uncertainties on all four SP parameters.\label{FIG3}}
\end{figure}

\section{Variation of stellar population parameters across and through the Fundamental Plane}

\subsection{Global trends with physical parameters}

To examine the trends of SP parameters in FP space, we first bin the galaxies along the $v_1-v_2-v_3$ coordinate axes.  That is, we set up bins in FP space with boundaries along those axes, of width 0.1 in the $v_1$ direction, 0.2 in the $v_2$ direction, and 0.2 in the $v_3$ direction.  The bins are narrower along $v_1$ because we wish to have a comparable number of bins along each dimension.  In each bin, we calculate the median value of each of the SP parameters.  After removing all bins with fewer than 5 galaxies, we are left with 92 bins.

We now consider two approaches to assessing trends in FP space, using the median values in each bin.  The first approach involves plotting each individual SP parameter against each FP parameter.  The problem with this type of analysis is that $r$, $s$, and $i$ are each correlated with one another, and it is difficult to determine to what extent a trend with a particular FP parameter is merely an artifact of that parameter's trend with {\it another} FP parameter.

Nevertheless, we have plotted the global variation of each SP parameter with each individual FP parameter in Figure 4.  In this figure, we show the median values of each of the four SP parameters vs. the corresponding $r$, $s$, and $i$ values of each bin for our dataset.  The figure includes a least squares fit to a linear trend for each of the individual SP-FP trends.  We also plot a dashed line representing a fit to a combination of ``directional derivatives'', that will be explained in Section 3.3.  The figure also shows the $R^2$ correlation coefficient for each plot.  All four SP parameters are seen to have a stronger correlation with velocity dispersion than with radius or surface brightness.

\begin{figure}
\centering
\includegraphics[width=0.45\textwidth]{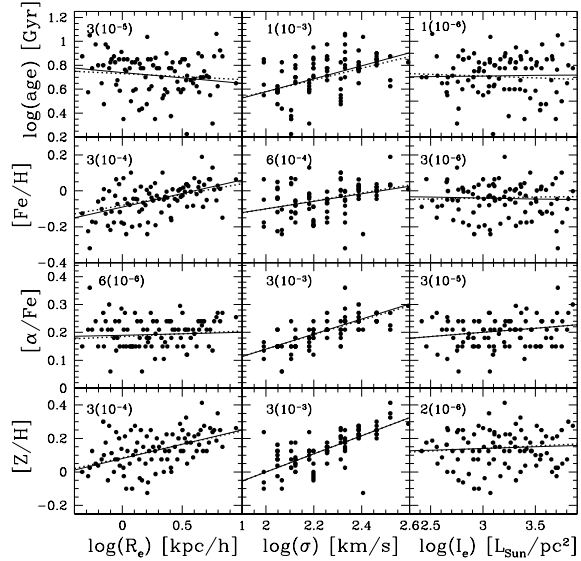}
\caption{Results of the stellar population modeling by bin in FP space.  Each point is the median value of one of the SP parameters in a bin in FP space, plotted against the corresponding $r$, $s$, or $i$ value at the center of that bin.  We also superimpose the best fit regression line to the plotted points ({\it solid line}) and a best fit line to a set of directional derivatives for $r$, $s$, and $i$ ({\it dashed line}, see Section 3.3).  The $R^2$ correlation coefficient is given in the upper left corner of each plot.  As \citep{proctor08b} shows, each of the possible SP values is quantized, though the quantization is most extreme for [$\alpha$/Fe].
\label{FIG4}}
\end{figure}

This figure can be compared with Figures 4-6 of \citetalias{graves09bb}.  (Note: We refer to [Mg/Fe] as [$\alpha$/Fe].  We also use [Z/H], while \citetalias{graves09bb} use [Mg/H], but the two quantities are nearly identical.)  Our results show agreement with \citetalias{graves09bb} and other authors, in finding a positive correlation between velocity dispersion and each of the SP parameters.  \citetalias{graves09bb} find a weak positive correlation between $r$ and both [Fe/H] and [Mg/H].  We find a similar positive correlation among these parameters (with [Z/H] in place of [Mg/H]), which may be slightly more pronounced in our data than in the \citetalias{graves09bb} data.  Additionally, \citetalias{graves09bb} claim a mild correlation between surface brightness and some of the SP parameters, but there are no such trends apparent in our Fig. 4.

\subsection{Variations along the axes of the 3D Gaussian}

As mentioned in the previous section, the correlations between $r$, $s$, and $i$ complicate the interpretation of Figure 4.  A more useful approach for displaying the data is to plot the full 3D distribution.  In Figures 5, 6, 7, and 8\footnote{These are interactive 3D figures, generated in the same manner as Figure 1.}, we show 3-dimensional FP space variation of $A=$log$(age)$, [Fe/H], [$\alpha$/Fe], and [Z/H] respectively.  When plotted this way, one can see more complex trends than those observed in Fig. 4.

\begin{figure}
\centering
\includegraphics[width=0.45\textwidth]{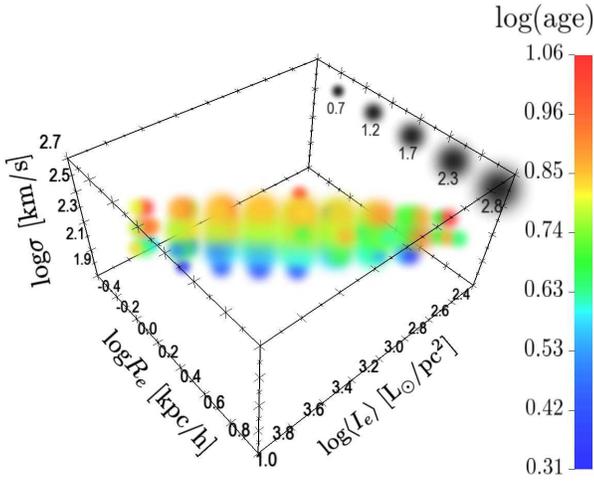}
\caption{Variation of log(age) across the Fundamental Plane, in 3D.  Each sphere represents a bin in FP space, including 5 or more galaxies.  The sphere is placed at the midpoint of the bin's $r-s-i$ coordinates, color-coded so that redder colors represent older ages, and bluer colors represent younger ages, as given by the color scale on the right of the plot.  The size of the sphere scales with the logarithm of the number of galaxies in the bin, as given by the scale established by the black spheres on the side of the plot.  The number labeling each of the black spheres is the logarithm of the number of galaxies in a bin represented by a sphere of that size.  As with Figure 1, readers using version 8.0 or higher of Adobe Reader can enable 3D interactive views of this plot by mouse clicking on the figure.\label{FIG5}}
\end{figure}


\begin{figure}
\centering
\includegraphics[width=0.45\textwidth]{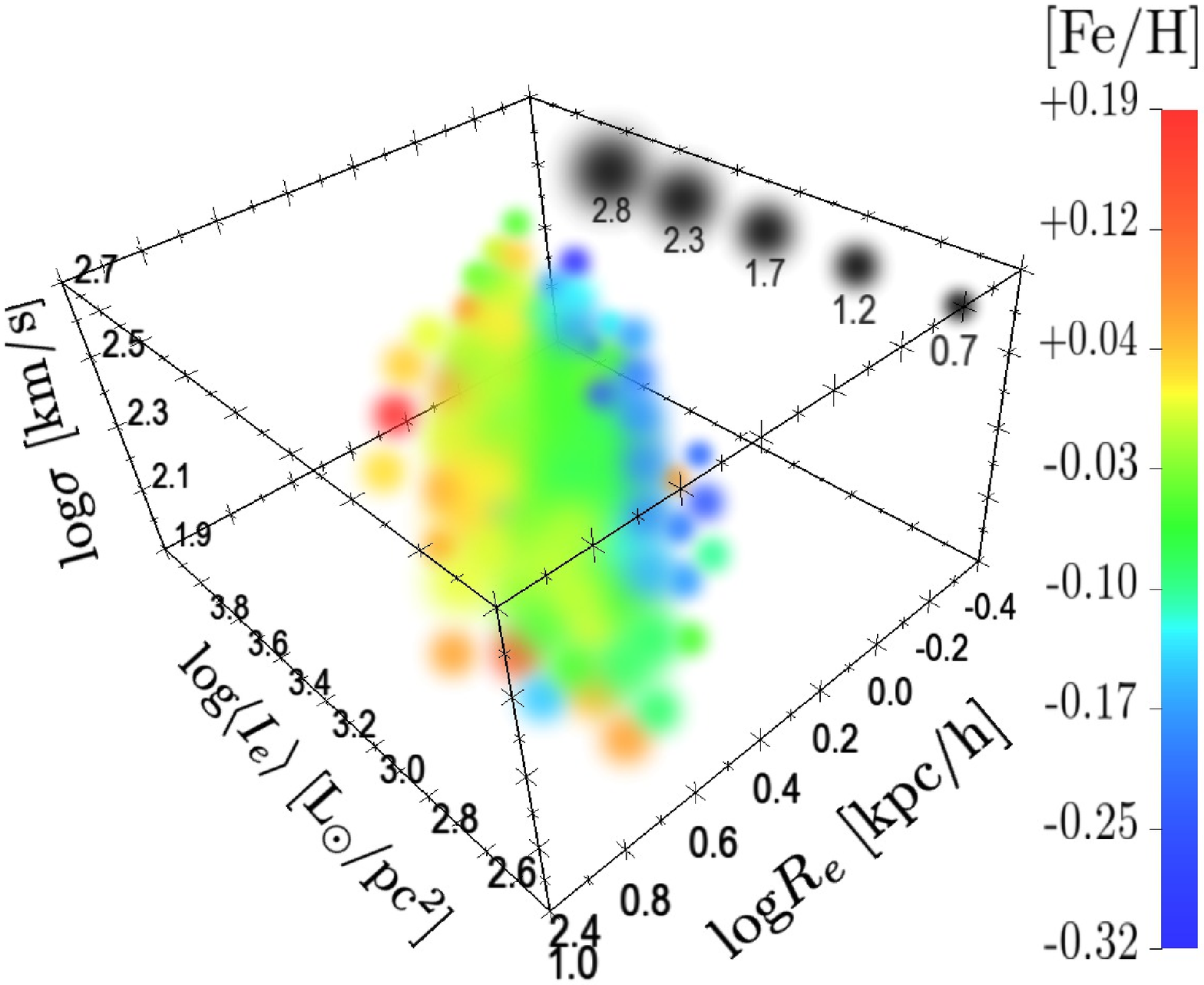}
\caption{Same as Figure 5, but with [Fe/H] rather than log(age).  Redder colors indicate higher values of [Fe/H].
\label{FIG6}}
\end{figure}

\begin{figure}
\centering
\includegraphics[width=0.45\textwidth]{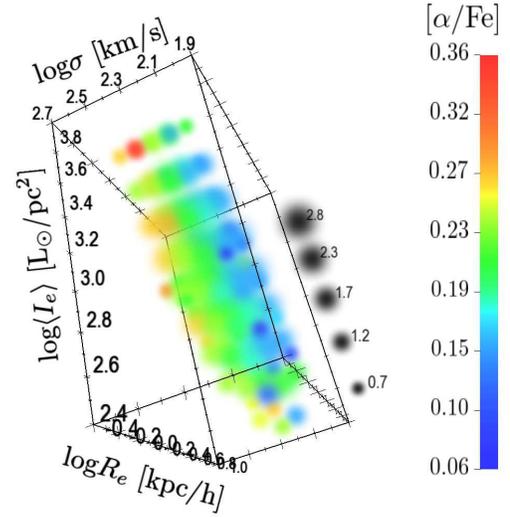}
\caption{Same as Figure 5, but with [$\alpha$/Fe] rather than log(age).  Redder colors indicate higher values of [$\alpha$/Fe].
\label{FIG7}}
\end{figure}

\begin{figure}
\centering
\includegraphics[width=0.45\textwidth]{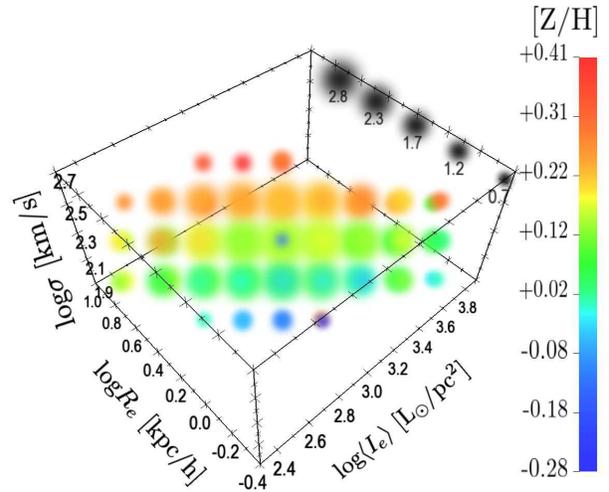}
\caption{Same as Figure 5, but with [Z/H] rather than log(age).  Redder colors indicate higher values of [Z/H].
\label{FIG8}}
\end{figure}

The second approach we take to assess the SP trends allows us to examine this complexity: We examine the trend of each SP parameter with each FP parameter {\it when the other two FP parameters are held constant}.  This approach makes use of the partial derivatives $\partial \mathcal{S} / \partial \mathcal{F}$, where $\mathcal{S}$ is used as a shorthand representation for any of the SP parameters and $\mathcal{F}$ is used as a shorthand representation for any of the FP parameters.

We calculate these partial derivatives in the following way.  First, we assume that the variation of each of the SP parameters in FP space can be fit by a least squares fit to a straight line.  We perform such a linear regression along the $v_1$, $v_2$, and $v_3$ directions individually, using the median values of each SP parameter in every bin for each fit.  For example, in fitting the variation of $A=$log$(age)$ along the $v_1$ direction, we fit the partial derivative of log$(age)$ with $v_1$ according to:

\be {\partial A \over \partial v_1} = {{N \Sigma v_{1i} A_i - \Sigma v_{1i} \Sigma A_i} \over {N \Sigma v_{1i}^2 - (\Sigma v_{1i})^2}}
\ee

\noindent where $v_{1i}$ and $A_{i}$ are, respectively, the $v_1$ position of the $ith$ bin and the median value of $A$ of the galaxies in the $ith$ bin.  The summation is performed over all $N=92$ bins containing 5 or more galaxies.  By then making the corresponding calculations for the $v_2$ and $v_3$ directions, we produce a vector of age variation in $v_1-v_2-v_3$ space.

While the components of this vector are partial derivatives, the vector itself can be thought of as the {\it gradient} of age in $v_1-v_2-v_3$ space, $\bigtriangledown A $.  In the next subsection, we generalize this approach to include the components of the gradient along directions other than $v_1$,$v_2$, and $v_3$.  In such cases, it no longer makes sense to describe the components of the gradient as merely `partial derivatives'.  Rather, we will refer to them as `directional derivatives'.  The directional derivative of $A$ with respect to $v_1$, which we write as $\bigtriangledown_{\hat{v}1}A$, is the change in $A$ along the $v_1$ direction, per unit change in $v_1$.  Likewise, $\bigtriangledown_{\hat{r}}A$ is the change in $A$ per unit $r$, and $\bigtriangledown_{\hat{m}}A$ the change in $A$ per unit $m = {\rm log}(mass)$.  

We have computed directional derivatives for each of the SP parameters: log($age$), [Fe/H], [$\alpha$/Fe], and [Z/H], which can be found in Table 2.  We also provide the statistical errors on each of the directional derivatives, $\epsilon$, as well as the absolute value of the ratio of each directional derivative to its own error, $\chi$ (e.g., for $\bigtriangledown_{\hat{\mathcal{F}}}A$, $\chi = |\bigtriangledown_{\hat{\mathcal{F}}}A| / \epsilon$).  Trends with significance $\chi > 3$ are bolded.


\begin{table*}
\centering
\begin{minipage}{150mm}
\begin{tabular}{@{}lrrrrrrrrrrrrrrr@{}}
\hline
FP parameter   &  $\bigtriangledown_{\hat{\mathcal{F}}}A$  &  $\epsilon$   &  $\chi$   & ~ & $\bigtriangledown_{\hat{\mathcal{F}}}[Fe/H]$   &  $\epsilon$   &  $\chi$ & ~ & $\bigtriangledown_{\hat{\mathcal{F}}}[\alpha /Fe]$   &  $\epsilon$   &  $\chi$  & ~ & $\bigtriangledown_{\hat{\mathcal{F}}}[Z/H]$   &  $\epsilon$   &  $\chi$
\\
\hline
\hline
$v_1$ & {\bf -1.47 }& {\bf 0.12} & {\bf 12.25}  & ~ & {\bf 0.37} & {\bf 0.10} & {\bf 3.70}  & ~ &  {\bf -0.24} & {\bf 0.05} & {\bf 4.80}  & ~ &  0.07 & 0.13 & 0.54 \\
$v_2$ & -0.04 & 0.04 & 1.00 & ~ & 0.05 & 0.02 & 2.50  & ~ &  -0.01 & 0.01 & 1.00 &  ~ & 0.05 & 0.03 & 1.67  \\
$v_3$ & 0.08 & 0.09 & 0.89  & ~ & {\bf 0.26} & {\bf 0.04} & {\bf 6.50}  & ~ &  {\bf 0.16} & {\bf 0.02} & {\bf 8.00} & ~ &  {\bf 0.46} & {\bf 0.04} & {\bf 11.50} \\
$r$ & {\bf -0.70} & {\bf 0.08} & {\bf 8.75} & ~  & {\bf 0.37} & {\bf 0.06} & {\bf 6.17}  &  ~ & -0.03 & 0.03 & 1.00 & ~ &  {\bf 0.32} & {\bf 0.07} & {\bf 4.57}  \\
$s$ & {\bf 1.16} & {\bf 0.11} & {\bf 10.55} & ~  & -0.11 & 0.08 & 1.38 &  ~ &  {\bf 0.29} & {\bf 0.04} & {\bf 7.25} & ~ &   0.25 & 0.10 & 2.50  \\
$i$ & {\bf -0.57} & {\bf 0.08} & {\bf 7.13} & ~  & {\bf 0.25} & {\bf 0.05} & {\bf 5.00}  &  ~ &  -0.02 & 0.03 & 0.67 & ~ &   {\bf 0.22} & {\bf 0.06} & {\bf 3.67}  \\
$m$ & {\bf 0.32} & {\bf 0.05} & {\bf 6.92} & ~  & 0.03 & 0.03 & 0.88   & ~ &   {\bf 0.11} & {\bf 0.02} & {\bf 6.44}  &  ~ &  {\bf 0.16} & {\bf 0.04} & {\bf 3.87} \\
$l$ & {\bf -0.39} & {\bf 0.04} & {\bf 11.01} & ~  & {\bf 0.20} & {\bf 0.03} & {\bf 7.62}  & ~ &   -0.02 & 0.01 & 1.19 & ~ &   {\bf 0.17} & {\bf 0.03} & {\bf 5.65}  \\
$m-l$ & {\bf 0.60} & {\bf 0.04} & {\bf 14.51} & ~  & {\bf -0.14} & {\bf 0.03} & {\bf 4.72} & ~ &   {\bf 0.11} & {\bf 0.02} & {\bf 6.96}  & ~ &   -0.01 & 0.04 & 0.18  \\
\hline
\end{tabular}
\caption{Stellar population trends in FP space.}
\end{minipage}
\end{table*}

In using this method, we have implicitly assumed that, for any given SP parameter, there is a direction in FP space along which that parameter increases linearly.  To test this hypothesis, we have also fit a quadratic curve to each of the SP trends along $v_1$, $v_2$, and $v_3$.  In every case, we find that the quadratic coefficient is consistent with zero, to within the statistical errors.  The assumption of linear variation in FP space thus seems justified.  In fact, even if there were minor deviations from linearity, our method would still be sufficient to illuminate these qualitative relationships between the SP and FP parameters.

One remarkable feature of the results in Tables 2 is the lack of variation of any of the SP parameters along the $v_2$ direction, the long dimension of the FP.  All of the SP parameters vary along a direction that is a superposition of the $v_1$ (`through the plane') and $v_3$ (`across the plane') axes.  Age variation is almost entirely through the plane, while [Z/H] variation is almost entirely across the plane, and [$\alpha$/Fe] and [Fe/H] are superpositions of the two.  The lack of variation of any of the SP parameters along $v_2$ is a major result of this paper, and is examined in more detail in Section 4.

We should also note here that $v_1$ is a quantity that has been studied by other authors.  It is simply the residual from the plane, measured along the direction orthogonal to the plane.  Several authors have examined correlations between various parameters and FP residual.  In some cases, this is measured as the residual along a different dimension, such as radius.  However, this scales with $v_1$ to within a constant scale factor, so increasing $v_1$ is proportional to increasing residual in $r$.  \citet{gargiulo09a}, for example, find an anticorrelation between age and residual in $r$, as well as an anticorrelation between [$\alpha$/Fe] and residual in $r$.  This is consistent with our finding of an anticorrelation between both of these quantities and $v_1$.  However, in contrast to \citet{gargiulo09a}, we do not find that the anticorrelation with [$\alpha$/Fe] is stronger than the one with age. 

\subsection{Variations with physical parameters}

As mentioned in Section 3.2, we wish to generalize the derivation of the directional derivative to directions other than $v_1$, $v_2$, and $v_3$.  We would like to examine how the SP parameter variations in $v_1-v_2-v_3$ space translate to variations in $r-s-i$ space.  In the case of the directional derivative with $v_1$, $\bigtriangledown_{\hat{v}1}A$ is exactly the same as $\partial A / \partial v_1$, as given by Equation 3.  While in the case of mass, the expression for the directional derivative of age with mass, $\bigtriangledown_{\hat{m}}A$, is a linear combination of partial derivatives with $v_1$, $v_2$, and $v_3$.  This is also the case for the directional derivative with luminosity, mass-to-light ratio, radius, velocity dispersion, and surface brightness.

Table 1 gives the $v_1$, $v_2$, and $v_3$ components of $r$, $s$, and $i$.  This provides the coordinate transformation from $\bigtriangledown_{\hat{v}1}\mathcal{S}$, $\bigtriangledown_{\hat{v}2}\mathcal{S}$, and $\bigtriangledown_{\hat{v}3}\mathcal{S}$ to $\bigtriangledown_{\hat{r}}\mathcal{S}$, $\bigtriangledown_{\hat{s}}\mathcal{S}$, and $\bigtriangledown_{\hat{i}}\mathcal{S}$.  We are also interested in calculating how each of the SP parameters varies with dynamical mass, luminosity, and mass-to-light ratio.  If we assume homology, then:

\be m = r + 2s + c_1
\ee

\be l = 2r + i + c_2
\ee

\noindent where $m = {\rm log}(mass)$, $l = {\rm log}(luminosity)$, and $c_1$ and $c_2$ are normalization constants.  Subtracting these equations, we express the logarithm of the mass-to-light ratio as

\be m-l = -r + 2s -i +c_1 -c_2
\ee

\noindent We then wish to derive $\bigtriangledown_{\hat{m}}\mathcal{S}$, $\bigtriangledown_{\hat{l}}\mathcal{S}$, and $\bigtriangledown_{m\hat{-}l}\mathcal{S}$.  As previously mentioned, we must be careful about how we define the directional derivative for a quantity such as mass, which does not represent any of the basis vectors in FP space.  We are defining $\bigtriangledown_{\hat{m}}\mathcal{S}$ to mean the change in an SP parameter $\mathcal{S}$ per unit $m$ along the gradient of $m$ in FP space, $\bigtriangledown m$.  This direction is $\bigtriangledown m = \hat{r}+2\hat{s}$, or the direction along which, for every increase ($\delta r$) in $r$ of one unit, there is a corresponding increase ($\delta s$) in $s$ of two units.  To normalize this vector to a change ($\delta m$) in $m$ of 1, we should actually divide by 5, because we require $\delta m = \delta r + 2 \delta s$, and $\delta s = 2\delta r$.  Thus, $\delta m = \delta r + 4\delta 4r = 5\delta r$, and $\delta r = 1/5$.

Thus, the directional derivative of the SP parameter $\mathcal{S}$, along the normalized gradient of $m$ is:

\be \bigtriangledown_{\hat{m}}\mathcal{S} = {1 \over 5} {\partial \mathcal{S} \over \partial r} + {2 \over 5} {\partial \mathcal{S} \over \partial s}
\ee

This is the change in $\mathcal{S}$ per unit change in $m$, provided that $m$ is changing along its gradient in $r-s-i$ space (along the direction $(\delta r, \delta s, \delta i) = (+1,+2,0)$).

We similarly derive:

\be \bigtriangledown_{\hat{l}}\mathcal{S} = {2 \over 5} {\partial \mathcal{S} \over \partial r} + {1 \over 5} {\partial \mathcal{S} \over \partial i}
\ee

\be \bigtriangledown_{m\hat{-}l}\mathcal{S} = -{1 \over 6} {\partial \mathcal{S} \over \partial r} + {1 \over 3} {\partial \mathcal{S} \over \partial s} -{1 \over 6} {\partial \mathcal{S} \over \partial i}
\ee

\noindent The resulting relationships between each SP parameter and each structural parameter are given in Table 2.  We also include statistical uncertainties $\epsilon_{\bigtriangledown S}$ and the ratio between $\bigtriangledown_{\hat{\mathcal{F}}}\mathcal{S}$ and $\epsilon_{\bigtriangledown S}$.  We note that the $m$, $l$, and $m-l$ directions in FP space that we have derived here correspond to the directions shown in Figure 1 (written as log($M$), log($L$), and log($M/L$) respectively).

We have also taken the individual directional derivatives $\bigtriangledown_{\hat{r}}\mathcal{S}$, $\bigtriangledown_{\hat{s}}\mathcal{S}$, and $\bigtriangledown_{\hat{i}}\mathcal{S}$, and computed the inferred variation of the SP parameters for a set of bins matching our bins' positions in FP space, then fit a regression line to those points.  The best fit lines are shown as the dashed lines in Figure 4.  The extremely close match between these best fit lines and the solid lines (from fits that do not assume there is a single direction in FP space along which the SP parameter varies linearly) offers further evidence that our linear fits are a good match to the real SP parameter variation.

As Table 2 shows, there are substantial differences between the different SP parameters in terms of how they vary with the FP parameters.  Three of the four SP parameters increase with increasing velocity dispersion.  The outlying case is [Fe/H], which shows no dependence on $s$ for fixed $r$ and $i$, despite the fact that, as discussed earlier, there is a global correlation between [Fe/H] and $s$ when $r$ and $i$ are allowed to vary.  Dependence on $r$ and $i$ varies from parameter to parameter, with [Z/H] and [Fe/H] increasing with increasing $r$ and $i$, age increasing with decreasing $r$ and $i$, and [$\alpha$/Fe] independent of $r$ and $i$.  Age varies most strongly with $v_1$, [Fe/H] most strongly with $v_1$ and $r$, [$\alpha$/Fe] most strongly with $s$, and [Z/H] most strongly with $v_3$.

We note that some of these trends have been identified by other authors as well.  For example, the relationship between age and FP residual, which we call $v_1$, was noted by \citet{forbes98x}.  And as we noted in the previous subsection, \citet{gargiulo09a} found relationships between FP residual and both age and [$\alpha$/Fe] that are consistent with ours, at least in terms of the sense of the correlation.  We also note that our estimate of $\partial A / \partial (m-l)$ is 0.60, which contrasts with $\sim 0.75$ (as estimated from \citealt{bruzual03x}, Figure 3), which is predicted by stellar population models.  The difference may well reflect differences between dynamical mass and stellar mass.

In summary, our results for the global trends of the SP parameters with respect to the FP parameters agree in broad terms with the trends observed by other authors, in that each SP parameter is seen to be positively correlated with velocity dispersion.  However, we have also taken the additional step of deriving the directional derivatives of each of the SP parameters with each of the FP parameters (Table 2).  That is, we have calculated the dependence of each SP parameter on each FP parameter by computing its gradient in FP space.  When analyzed in this way, it is no longer the case that all four SP parameters depend more strongly on velocity dispersion than any other FP parameter.  Most interestingly, the vectors along which the SP parameters vary are closely aligned with the axes of the 3D Gaussian that defines the FP, with age varying almost entirely with $v_1$, [Z/H] varying almost entirely with $v_3$, and none of the SP parameters varying along $v_2$.

\subsection{Comparing `slices' of 6dFGS FP with those of SDSS}

We have shown how examining the full 3-dimensional distribution of stellar population variations in FP space gives one a clearer picture of the SP trends than one would get from collapsing the trends down to a 2-dimensional distribution.  We now present another set of plots that we can compare to the SDSS results presented in \citetalias{graves09bb}.  \citetalias{graves09bb} Figures 7-10 show the variation of the SP parameters within a given slice of the FP: below the plane, within the plane, and above the plane.  We produce similar figures for our data in our Figures 9-12.

\begin{figure*}
\centering
\begin{minipage}{150mm}
\includegraphics[width=1.0\textwidth]{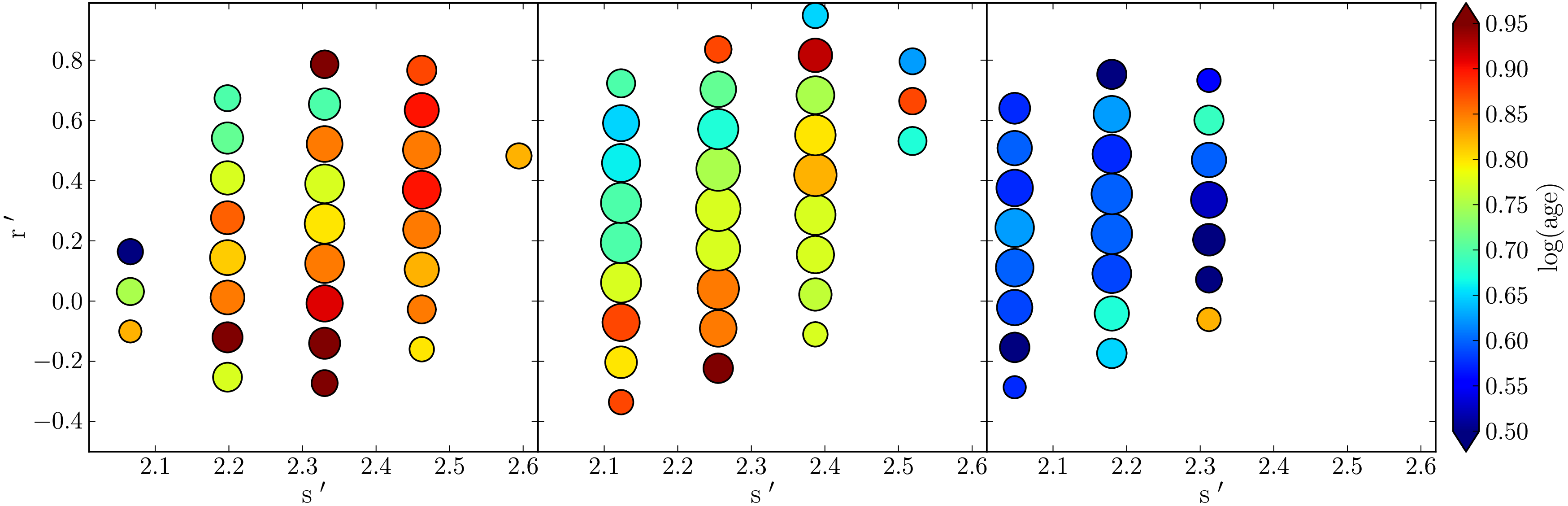}
\caption{Each panel shows the distribution of bins (described in Section 3.1) across $r'$ and $s'$ for a different slice of the FP.  The left panel is the $v_1<0$ slice (`below the plane'), the middle panel is $v_1\sim 0$ (`the midplane'), and the right panel is the $v_1>0$ slice (`above the plane').  $r'$ and $s'$ are as described in Section 3.4: $r'$ is the value of $r$ within the plane for fixed $s$, and $s'$ is the value of $s$ within the plane for fixed $r$.  The area of each circle is proportional to the logarithm of the number of galaxies in the bin.  The color of each circle represents the median value of log($age$) in the bin, as given by the color scale shown on the right.  Redder colors correspond to older ages, and bluer colors correspond to younger ages.\label{FIG9}}
\end{minipage}
\end{figure*}

\begin{figure*}
\centering
\begin{minipage}{150mm}
\includegraphics[width=1.0\textwidth]{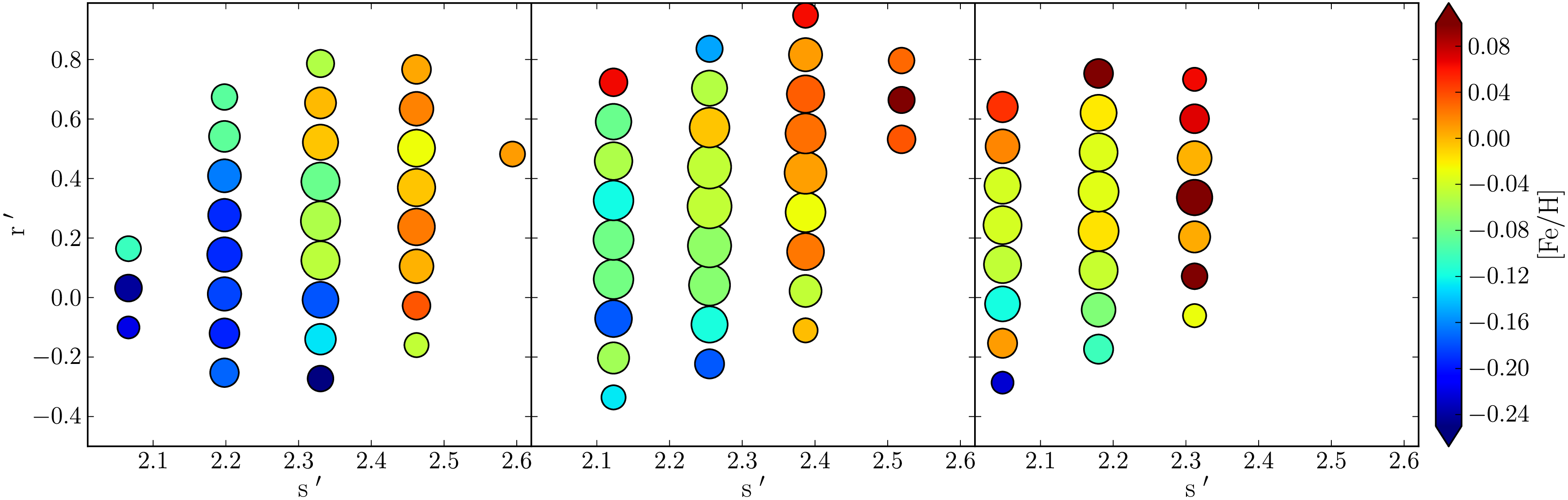}
\caption{Same as Figure 9, but for [Fe/H] rather than age.  Redder colors correspond to higher values of [Fe/H].\label{FIG10}}
\end{minipage}
\end{figure*}

\begin{figure*}
\centering
\begin{minipage}{150mm}
\includegraphics[width=1.0\textwidth]{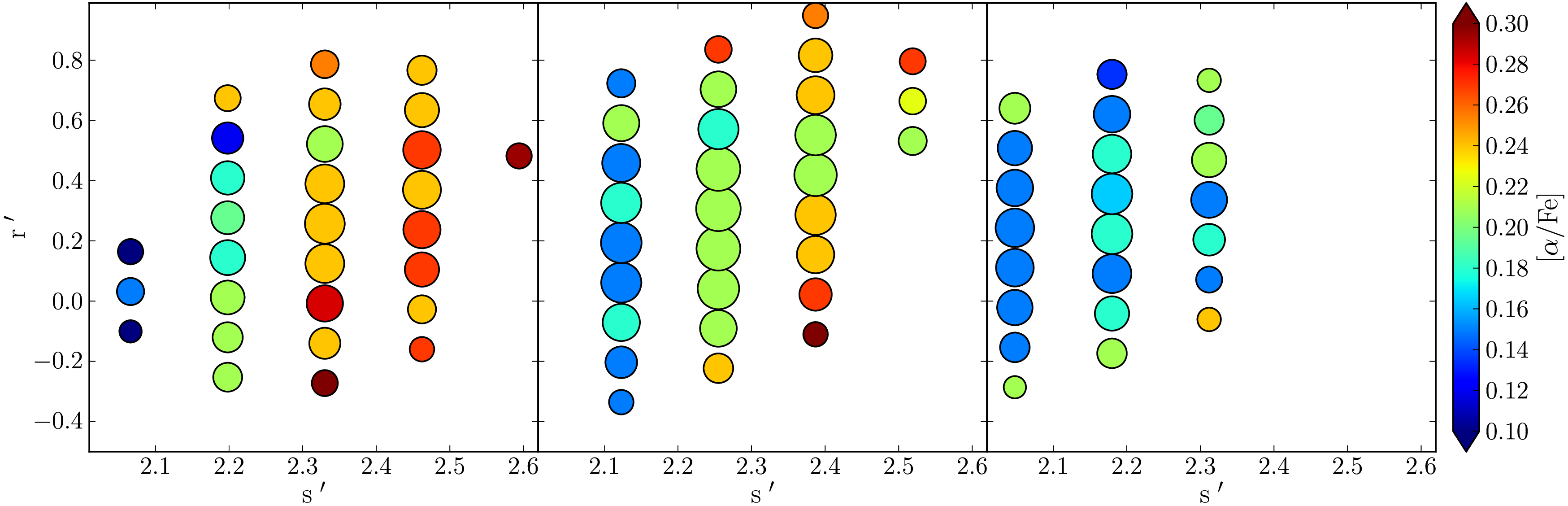}
\caption{Same as Figure 9, but for [$\alpha$/Fe] rather than age.  Redder colors correspond to higher values of [$\alpha$/Fe].\label{FIG11}}
\end{minipage}
\end{figure*}

\begin{figure*}
\centering
\begin{minipage}{150mm}
\includegraphics[width=1.0\textwidth]{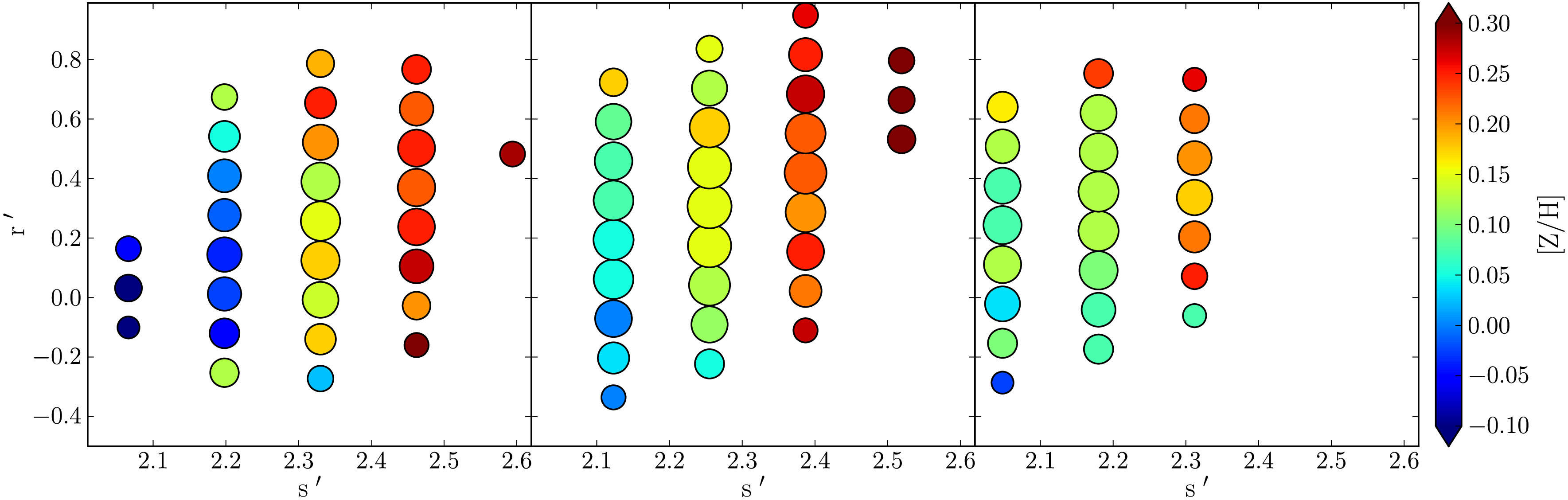}
\caption{Same as Figure 9, but for [Z/H] rather than age.  Redder colors correspond to higher values of [Z/H].\label{FIG12}}
\end{minipage}
\end{figure*}

Figure 9 shows our FP bins distributed in three slices.  The left panel shows the distribution of bins directly ``below'' the plane.  That is, it shows the bins with central $v_1$ values of -0.1, equivalent to negative $\Delta i$ in the nomenclature of \citetalias{graves09bb}.  The two axes are $r'$ and $s'$, where $r'$ is the direction {\it within} the FP along which $r$ increases but $s$ remains constant and $s'$ is the direction within the FP along which $s$ increases but $r$ remains constant.  The middle panel is the equivalent plot for the `midplane', which includes the bins with central $v_1=0$, and the right panel is the equivalent plot for bins with central $v_1=0.1$, equivalent to positive $\Delta i$ in the \citetalias{graves09bb} nomenclature.  Our binning scheme also includes a few bins with central $v_1=-0.2$ and central $v_1=0.2$, but they include very few galaxies, and are not plotted here.

In each plot, the area of each circle is proportional to the logarithm of the number of galaxies in the bin, and the color of the circle scales with log($age$), with redder points corresponding to older galaxies.  The main trend seen here is the large variation occurring between different slices of $v_1$, with the bins `above' the plane including more young galaxies.

In Figure 10, we present the equivalent plot for [Fe/H].  In this case, [Fe/H] increases with increasing $v_1$.  We also observe that [Fe/H] increases with increasing $r'$ and $s'$.  In Figure 11, we present the equivalent plot for [$\alpha$/Fe].  [$\alpha$/Fe] is positively correlated with $s'$ at low $v_1$, whereas [$\alpha$/Fe] is more consistently low and only weakly dependent on $s'$ at high $v_1$.  In Figure 12, we present the equivalent plot for [Z/H]. [Z/H] increases sharply with increasing $s'$, though the trend is somewhat muted for the high $v_1$ slice relative to the other two slices.

Each of these trends closely tracks the SP trends that \citetalias{graves09bb} observe in the SDSS Fundamental Plane, when the data are plotted in the same manner (their Fig. 7-10).  (We do note just two differences: \citetalias{graves09bb} find a much stronger trend between age and $s'$ and a weaker trend between [Fe/H] and $r'$ than we do.)  This suggests that the underlying trends in the SDSS and 6dFGS datasets are very nearly the same.

\section{Discussion}

\citet{graves10y} characterize the SP parameter variation in FP space as `the 2D family of early-type galaxy stellar populations and their structural properties'.  This is contrasted with the 1D mass sequence of galaxies, which was the focus of earlier studies.  Earlier work, such as \citet{nelan05x} and \citet{thomas05a}, examined the variation of the SP parameters with $\sigma$, which was used as a proxy for galaxy mass.  To first order, one can imagine these 1D relationships as variations along the $\sigma$ projection within the FP.

Graves, Faber, \& Schiavon improve on this analysis by also exploring SP variations {\it through} the plane.  The second dimension in the 2D family of early-type SPs represents residuals from the FP.  SP variations along this dimension can be interpreted as evolutionary differences, with aging stellar populations evolving in mass-to-light ratio.  Our results show, for example, that age varies more strongly than the other SP parameters with FP residual.  As \citet{graves10x} point out, however, the thickness of the plane may be better explained by genuine structural differences than by the fading of stellar populations.

Table 3 of \citet{graves10x} summarizes the qualitative relationships between each SP parameter and each FP parameter for their dataset.  However, as explained in the previous section, the three directions in FP space that the authors consider are not orthogonal.  They are what we refer to as $r'$, $s'$, and $\Delta i$, where $r'$ and $s'$ are as defined in the previous section (the direction within the plane along which $r$ increases but $s$ remains constant and the direction within the FP along which $s$ increases but $r$ remains constant, respectively) and $\Delta i$ is the residual from the plane along the $i$ direction.

This scheme is similar to, but not quite the same as, our orthogonal $v_1-v_2-v_3$ basis set.  $r'$ is exactly identical to $v_2$.  $\Delta i$ is defined somewhat differently from $v_1$, but the two are equivalent to one another to within a constant scaling factor.  The one dimension that Graves \etal ~use that is different from any of our basis vectors is $s'$.  It includes a significant $v_3$ component, but also has a $v_2$ component.  However, since none of the SP parameters varies with $v_2$, we would expect that any SP trend with $v_3$ would also be apparent in $s'$.

And indeed, if one compares \citet{graves10x} Table 3 with Table 2 from this paper, one gets remarkable agreement, from substituting $\Delta i$ for $v_1$, $r'$ for $v_2$, $s'$ for $v_3$, [$\alpha$/Fe] for [Mg/Fe], and [Z/H] for [Mg/H].  If one classifies any trend detected in our data at a significance of less than $3\sigma$ as `null' in the nomenclature of Graves \& Faber, then the only differences between the two tables upon making such substitutions are that we find no statistically significant trends between age and $v_3$, nor between [Z/H] and $v_1$.  So despite the differences in the binning scheme and despite the fact that Graves \etal ~stack the spectra in each bin, whereas we derive SP parameters for individual galaxies, we find broadly similar trends for each of the SP parameters.

While the analysis by \citet{graves10x} focuses on the `2D' variation of SP parameters, we would actually like to focus in particular on the third dimension in FP space, the dimension along which none of the SP parameters seems to vary.  As we show in Table 2, the vectors along which the SP parameters vary seem to be more closely aligned with the $v_1$ and $v_3$ vectors than they do with the vectors of any simple physical quantity.  Age varies almost purely along $v_1$, [Z/H] varies almost purely along $v_3$, and [Fe/H] and [$\alpha$/Fe] vary along superpositions of $v_1$ and $v_3$, with no component in $v_2$.

This is a remarkable result, as we had no physical motivation in choosing the directions of these vectors, allowing the data to determine the axes of the fitted 3D Gaussian.  However, because the SP parameters are so closely aligned with the axes of the 3D Gaussian, it now seems likely that there is in fact some physically meaningful reason for the $v_2$ and $v_3$ axes to be oriented in these particular directions in FP space.  What we require is a hypothesis that explains both the distribution of galaxies in FP space, and why the SP parameters vary along the 3D Gaussian.

\subsection{$v_2$ and merging history}

In Section 5 of \citetalias{graves09bb}, the authors discuss the question of why the SP parameters apparently vary with velocity dispersion, but not radius.  They suggest that, as seen in {\it N}-body simulations \citep{robertson06a,boylankolchin05bb}, galaxies of comparable mass and luminosity with different merger histories can have wildly different radii and surface brightnesses.  Velocity dispersion, on the other hand, is relatively independent of merger history.  The argument is that if SP parameters are independent of merger history, then it would follow that SP parameters would vary with velocity dispersion, but not radius.

In this work, however, we find that there is in fact variation of SP parameters with radius, even within the plane, as most of the SP parameters vary with $v_3$, which includes an $r$ component.  We believe our results can be explained by a somewhat modified version of the \citetalias{graves09bb} hypothesis: The variations in merger history add significant scatter to the correlations between radius and the SP parameters, but they do not eliminate them completely.

This idea is supported by the simulations of \citet{kobayashi05cc}.  Figure 5 of that paper shows how variable merger histories have little impact on the scaling relations of $\sigma$ with other physical parameters of the galaxy, while adding significant scatter to scaling relations with $r$.  However, despite the significantly added scatter, a residual correlation between $r$ and the other galaxy parameters remains.  There is no reason why this should not also be the case with respect to correlations between $r$ and the SP parameters.

We then offer the following scenario: As one moves along the $v_3$ axis towards increasing $r$, $s$, and $i$, one finds galaxies of increasing mass, luminosity, and metallicity.  (This is because larger $r$, $s$, and $i$ necessarily implies larger mass and luminosity, as $log(mass)=r+2s$ and $log(luminosity)=i+2r$.)  This could simply be driven by the increasing total mass of the system, or the increasing mass of the dark matter halo that seeded the galaxy.  However, as one moves along the $v_2$ axis, one finds galaxies of increasing radius and decreasing surface brightness associated with variations in merger history.  Indeed, if the FP were precisely the virial plane, then $v_2$ would be precisely $luminosity/radius^3$ (luminosity density).  We conclude then that luminosity density is determined by, or at least heavily influenced by, merger history, which in turn has little to no impact on the galaxy's stellar population.

This hypothesis is also consistent with the results of \citet{trujillo11}, who examine samples of elliptical galaxies at both $z\sim 0$ and $z\sim 1$, and find that evolution in size is independent of stellar age.  The authors suggest that this argues in favor of size evolution since $z\sim 1$ being driven more by dry mergers than by a ``puffing up'' scenario \citep{fan08gg,fan10a,damjanov09b} in which growth in galaxy radius is due to the expulsion of gas by AGN or supernova-driven winds.

If our hypothesis is correct, then galaxies are spread along the $v_2$ and $v_3$ directions by physically unrelated processes, related to secular migration and the galaxy's initial conditions respectively.  It is then something of a `coincidence' that these two processes distribute the galaxies along orthogonal directions in FP space.  Of course, we cannot rule out small deviations from orthogonality that our fitting method is unable to detect.

\subsection{Simulations of merger history variations across the Fundamental Plane}

One approach we can take to investigate this issue further is to examine the distribution of galaxies in FP space, as seen in simulations, for which the merger history of each galaxy is known.  To that end, we have examined the simulations presented in \citet{kobayashi05cc}.  In that paper, the author simulates the formation and evolution of 128 galaxies using a smoothed particle hydrodynamics method and a special purpose computer GRAPE (GRAvity PipE).  Because both the FP parameters and formation mechanisms of each of these simulated galaxies is known, we can compare the merger histories to the galaxy distribution in FP space, and determine what impact merger history has on a galaxy's position in FP space.  We describe this investigation below.

The details of the GRAPE-SPH code are presented in \citet{kobayashi04aa}, but we briefly recount the main features of the simulation here:  The initial conditions are 74 spherical regions with Cold Dark Matter initial fluctuations, which produces 83 elliptical galaxies and 45 dwarf galaxies.  As well as the kinematics of the dark matter and gas particles, the relevant baryon physics (i.e., radiative cooling, star formation, chemical enrichment, and supernova feedback) are included.  Computing the particle distribution from $z\sim 25$ to $z=0$, the time evolution of the internal structures of galaxies are predicted.

Physical parameters are measured for each of these simulated galaxies, and the scaling relations are examined in \citet{kobayashi05cc}.  As Figure 8 of that paper shows, if one plots the resulting parameters in $\kappa$-space \citep{bender92z}, it appears that divergent merging histories increase the scatter of the Fundamental Plane.

We note, however, that the $\kappa$-space representation does not account for the tilt of the FP.  In Figure 13, we thus produce our own plot of the ``edge-on'' view of the FP for these simulated galaxies in FP space.  We have fit the FP for the simulated galaxies in question, and find that they follow an FP relation of $r = 1.30s-0.54i-1.03$.  Deviations from the line in Figure 13 thus represent deviations from the FP.  Galaxies with all of the realized merging histories appear to follow the same FP relation.

\begin{figure}
\centering
\includegraphics[width=0.45\textwidth]{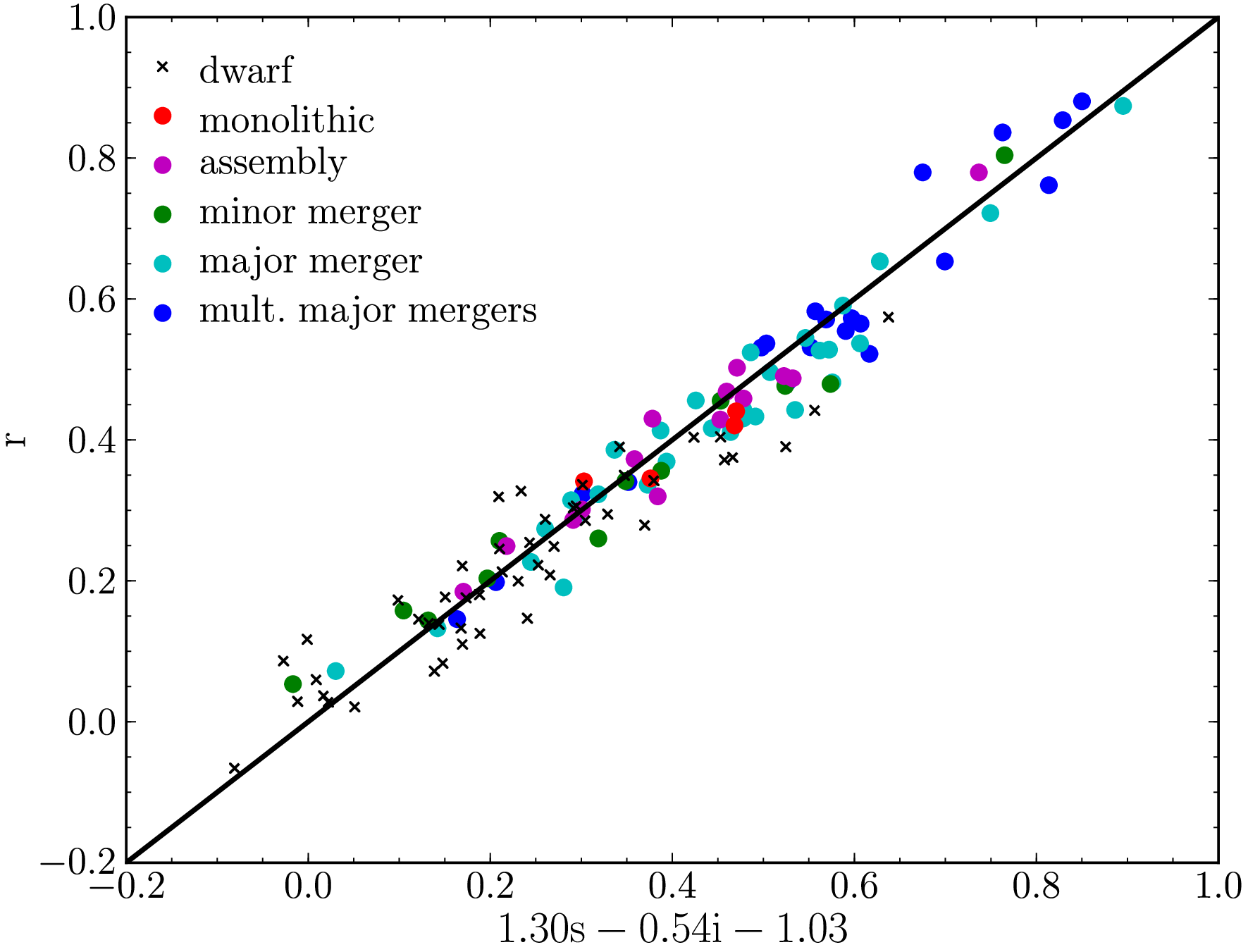}
\caption{Fundamental plane relation for galaxies in the \citet{kobayashi05cc} simulation.  The colors show the merging histories of the galaxies, following the classification scheme explained in Section 3.1 of that paper: [E1] monolithic (red), [E2] assembly (magenta), [E3] minor merger (green), [E4] major merger (cyan), [E5] multiple major mergers (blue), and [D1-5] dwarfs (black crosses).  The plot shows that while there is variation in merger history along the plane, there is no apparent trend between merger history and scatter off of the plane.\label{FIG13}}
\end{figure}

In Figure 14, we plot $s$ vs $v_2$, where $v_2 = 0.475r-0.880i$, as appropriate for the FP relation for the simulated dataset in question.  Dwarf galaxies are excluded, as they nearly all have $s<2.0$, and thus do not have counterparts in the 6dFGSv sample.  This plot shows that there are clear correlations between merger history and position in FP space.  The ellipticals that formed via monolithic collapse or the assembly of subgalaxies are preferentially found to have large velocity dispersions, and {\it low} values of $v_2$, which corresponds to small radii and large surface brightnesses.  For any fixed value of $s$, there appears to be a relationship between the value of $v_2$ and the merger history of the galaxy, such that galaxies with large radii and small surface brightnesses are likely to have been formed by one or more major mergers.

\begin{figure}
\centering
\includegraphics[width=0.45\textwidth]{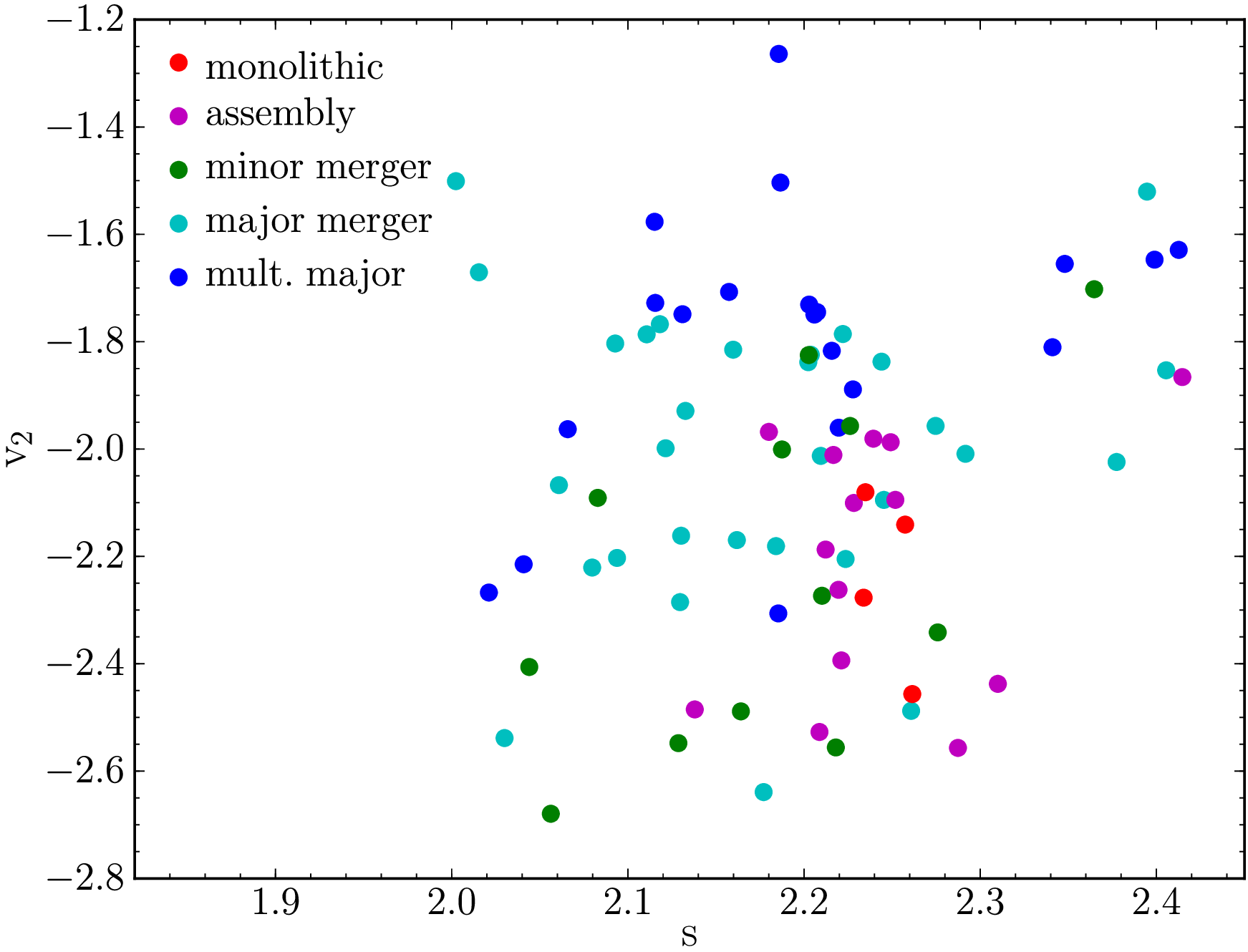}
\caption{$v_2$ vs. $s$ for the elliptical galaxies in the \citet{kobayashi05cc} simulation.  The colors are defined the same as in Figure 13.  We omit dwarf galaxies from the plot, as they would not be included in 6dFGSv anyway.  The plot shows clear trends between merger history and both $v_2$ and $\sigma$.\label{FIG14}}
\end{figure}

In Section 4.1, we hypothesized that variations in merger history elongate the FP along the $v_2$ direction.  Figure 14 would seem to confirm that this is the case for the simulations, as we do indeed see that for fixed $s$, there is a dependence of $v_2$ on merger history.

There is, however, one difference between these simulations and the real data of 6dFGSv that complicates our interpretation, and should be noted.  The value of $b$ in these simulations is -0.54, very close to the -0.5 value that corresponds to constant luminosity along the $v_2$ direction, and quite different from the virial expectation of -1.  In the simulations, when one compares different galaxies at the same velocity dispersion, but different values of $v_2$, the luminosities are nearly the same.  Thus, the variations in merger history do not lead to variations in luminosity, when one controls for $s$.  This does not hold in the real data however.  Since the 6dFGSv sample has $b=-0.885$, the galaxies with larger $v_2$ have not just larger radii, but larger luminosities.

Likewise, as discussed in \citetalias{magoulas11x}, essentially all authors have found a value of $b$ in the range $-0.9<b<-0.7$, regardless of waveband.  The lower right panel of \citet{kobayashi05cc} Figure 5 confirms the mismatch in the slopes of $r$ vs. $i$ between the simulations and real data.  The reason for this discrepancy is unclear.  It is possible that this is because the secondary starbursts are rarely induced by mergers in the simulations.  If secondary starbursts occur, the metallicity should be larger, which may break the mass-metallicity relation of galaxies.  This should be addressed in future simulations with higher resolutions.

Since we do not find $b=-0.5$ in our dataset, there remains something of a problem with the hypothesis that $v_2$ represents variations in luminosity density from divergent merger histories.  What, precisely, does it mean for two galaxies to be ``the same except for merger history''?  In the simulations, luminosity is held nearly constant along $v_2$.  In the real data, however, what is held constant along $v_2$?  Neither mass nor luminosity is fixed along $v_2$, though they both vary far more slowly along $v_2$ than along $v_3$.  One possibility is that galaxies along $v_2$ (with fixed $v_1$ and $v_3$) had the same initial mass at an earlier epoch, but have accreted different amounts of additional material.  Since the slope of $v_2$ also does not match the virial expectation of $b=-1$, galaxies with larger $v_2$ not only have larger radii, but slightly larger mass-to-light ratios.  We would thus conclude that galaxies which underwent major mergers would, on average, have somewhat greater mass-to-light ratios.

\section{Conclusions}

We have presented stellar population parameters for 7132 early-type galaxies and spiral bulges in the 6dFGSv survey, derived using Lick indices, following the procedure described in \citet{proctor08b}.  We have binned the galaxies in FP space, and fit the vectors along which each SP parameter varies.  Each of the SP parameters appears to vary with some or all of the structural parameters of the FP (effective radius, velocity dispersion, and surface brightness).  However, for log($age$) and [Z/H], the variation is more closely aligned with the axes of the 3D Gaussian to which the FP has been fit than with any physical parameter.  Age varies almost entirely in the $v_1$ direction (through the plane), [Z/H] varies almost entirely in the $v_3$ direction (across the plane), while [Fe/H] and [Mg/Fe] vary along both $v_1$ and $v_3$.  None of the SP parameters varies in the $v_2$ direction (along the plane).  The components of each of these vectors are given in Table 2.

These trends are similar to those seen in SDSS data by \citet{graves09bb} (GFS), though we have a somewhat different interpretation.  \citetalias{graves09bb} find weaker trends of the SP parameters with radius, though we argue that this is due to the fact that the orientation of their bins is not orthogonal.  We suggest that the axes of the 3D Gaussian to which the FP has been fit may in fact have some fundamental physical meaning.  Our hypothesis is a modified version of that suggested by \citetalias{graves09bb}: The $v_3$ direction represents a mass sequence, while the $v_2$ direction represents a variation in luminosity density caused by variations in merger history, which would be disconnected from SP effects.  Neither mass nor luminosity remains constant as one moves along the $v_2$ direction though, so the precise definition of ``variable merger history'' is still somewhat ambiguous.  This interpretation is supported by the {\it N}-body simulations of \citet{kobayashi04aa}, which shows a clear variation in merging history along $v_2$.  We also note that the simulations show no apparent correlation between merger history and residual from the FP.

Finally, we note that the fact that much of the SP variation (particularly that of age) is {\it through} the plane, it may be possible to account for SP variation in the derivation of distances, and reduce the distance errors on each galaxy.  This possibility will be explored in an upcoming paper that will present the 6dFGSv galaxy distances and peculiar velocities.

\vspace{10 mm}

Three-dimensional visualisation was conducted with the S2PLOT progamming library \citep{barnes06x}.  We wish to thank Chris Fluke for his help with preparing these plots.  We also thank Alex Merson for providing the 6dFGS group catalog, and Genevieve Graves and Max Spolaor for helpful comments and discussion.  We acknowledge the efforts of the staff of the Australian Astronomical Observatory (AAO), who have undertaken the observations and developed the 6dF instrument.

We acknowledge support from Australian Research Council (ARC) Discovery Ð Projects Grant (DP-0208876), administered by the Australian National University.  CM and JM acknowledge support from ARC Discovery Ð Projects Grant (DP-0208876).  CM is also supported by a scholarship from the AAO.  JM thanks the University of Melbourne for a professorial fellowship, and the Australian Astronomical Observatory for a Distinguished Visitorship.





\vfill
\end{document}